OPEN

# Spectroscopic evidence of odd frequency superconducting order

Avradeep Pal[1], J. A. Ouassou[2], M. Eschrig[3], J. Linder[2] & M. G. Blamire[1]

Received: 11 October 2016
Accepted: 30 November 2016
Published: 20 January 2017

**Spin filter superconducting S/I/N tunnel junctions (NbN/GdN/TiN) show a robust and pronounced Zero Bias Conductance Peak (ZBCP) at low temperatures, the magnitude of which is several times the normal state conductance of the junction. Such a conductance anomaly is representative of unconventional superconductivity and is interpreted as a direct signature of an odd frequency superconducting order.**

In the context of paired electrons in superconductors (S), Pauli exclusion requires the electron pair to be odd in exchange in either spin, momentum, time (frequency) or across all three parameters. Spin singlet (odd spin) Cooper pairs are the standard carriers in conventional superconductors. Although much rarer, there is strong evidence for odd momentum (odd parity) superconductivity in $Sr_2RuO_4$ and in UPt[1–3] and in artificially engineered hybrid structures[4]. So far, only indirect[5,6] or weak[7] signatures of odd frequency superconductivity has been obtained at S/F interfaces.

It is now widely accepted that odd *frequency* Cooper pairs can be generated at the interface of superconductors and ferromagnets. Where there is a region of inhomogeneous magnetization[8–10], such pairs acquire a net spin and hence are immune to pair breaking due to the internal exchange field of the ferromagnet and can traverse distances much longer than the relevant singlet coherence length. Strong, but indirect evidence for the existence of such superconductivity has mainly been obtained through indirect measurements such as a long-ranged supercurrent[5,6,11] or proximity effect[12].

Odd frequency Cooper-pairing should also give rise to an unconventional (non-BCS like) density of states (DOS) in the superconductor with an enhanced DOS at the Fermi level[13–15]. Evidence for such a DOS has recently been obtained via scanning tunnelling measurements of the local DOS of a Nb film proximity-coupled to a diffusive Ho layer[7] but, because this has to be performed on an interface remote from that which is generating the odd-frequency pairing, only localized weak spectroscopic signatures of odd frequency pairing were detected.

In parallel with the work on diffusive ferromagnets, superconducting tunnel junctions with ferromagnetic insulator (FI) barriers have recently shown a range of intriguing effects, such as the appearance of a Josephson current with an unconventional pure second harmonic current-phase relation independent of a $0-\pi$ transition[16], and an interfacial exchange field in the S layer[17], which suggest that odd-frequency pairing is also being generated at the S/FI interface. Indeed, it has been very recently suggested in a theoretical study[18] that strong odd-frequency pairing exists in Meservey-Tedrow type experiments with FIs[19] that show Zeeman split DOS.

In this work, we present differential conductance measurements of NbN/GdN/TiN tunnel junctions, where GdN serves the purpose of both a spin active interface as well as a tunnel barrier - enabling direct measurement of the spatially averaged tunnelling DOS at the S/FI interface. The measured ZBCPs in such differential conductance measurements are larger by at least an order of magnitude than reported for diffusive systems, and hence provide definitive evidence for an odd frequency superconductivity at S/FI interfaces.

TiN was chosen as the metallic layer because an all nitride stack is required for the stability of GdN. Moreover, as stated later in this paper, one of the theoretical requirements for the observation of ZBCPs necessitates the choice of a metal, which has a Fermi vector largely different from superconducting NbN. Other possible metallic candidates like Au, Al and Cu have comparable Fermi vector as that of NbN.

## Results and Discussion

**Expreimental.** In Fig. 1, we show the temperature dependence of resistance of a 3 nm GdN junction. The spin-filter effect is clearly visible from the decrease in junction resistance below 33 K as exchange splitting of the conduction band lowers the transmission probability of one spin channel in comparison to the other[20]. A sharp drop in resistance is observed below 14 K due to the superconducting transition of the NbN layer. Below 14 K, the observed rise in low bias resistance is due to the opening of the NbN gap and freezing out of sub-gap conductance. This rise of resistance below 14 K is reflective of a decreasing sub-gap resistance $R_S$ to normal state resistance $R_n$ ratio at low temperatures, and is therefore a signature of good quality junctions[21]. The drop in resistance below 4 K is due to the evolution of a zero bias conductance peak. To confirm the non-superconducting nature of the TiN used in these experiments, the temperature dependence of the resistance of un-patterned films of TiN/GdN grown in the same deposition run as that of the junctions, is shown in the bottom inset to Fig. 1. This shows no detectable superconducting transition above 1.6 K (the temperature limit of the cryostat used for Resistance vs Temperature (RT) measurements).

[1]Department of Materials Science, University of Cambridge, 27 Charles Babbage Road, Cambridge CB3 0FS, United Kingdom. [2]Department of Physics, NTNU, Norwegian University, N-7491 Trondheim, Norway. [3]SEPnet and Hubbard Theory Consortium, Department of Physics, Royal Holloway, University of London, Egham, Surrey TW20 0EX, United Kingdom. Correspondence and requests for materials should be addressed to A.P. (email: avradeep@gmail.com) or J.L. (email: jacob.linder@ntnu.no)

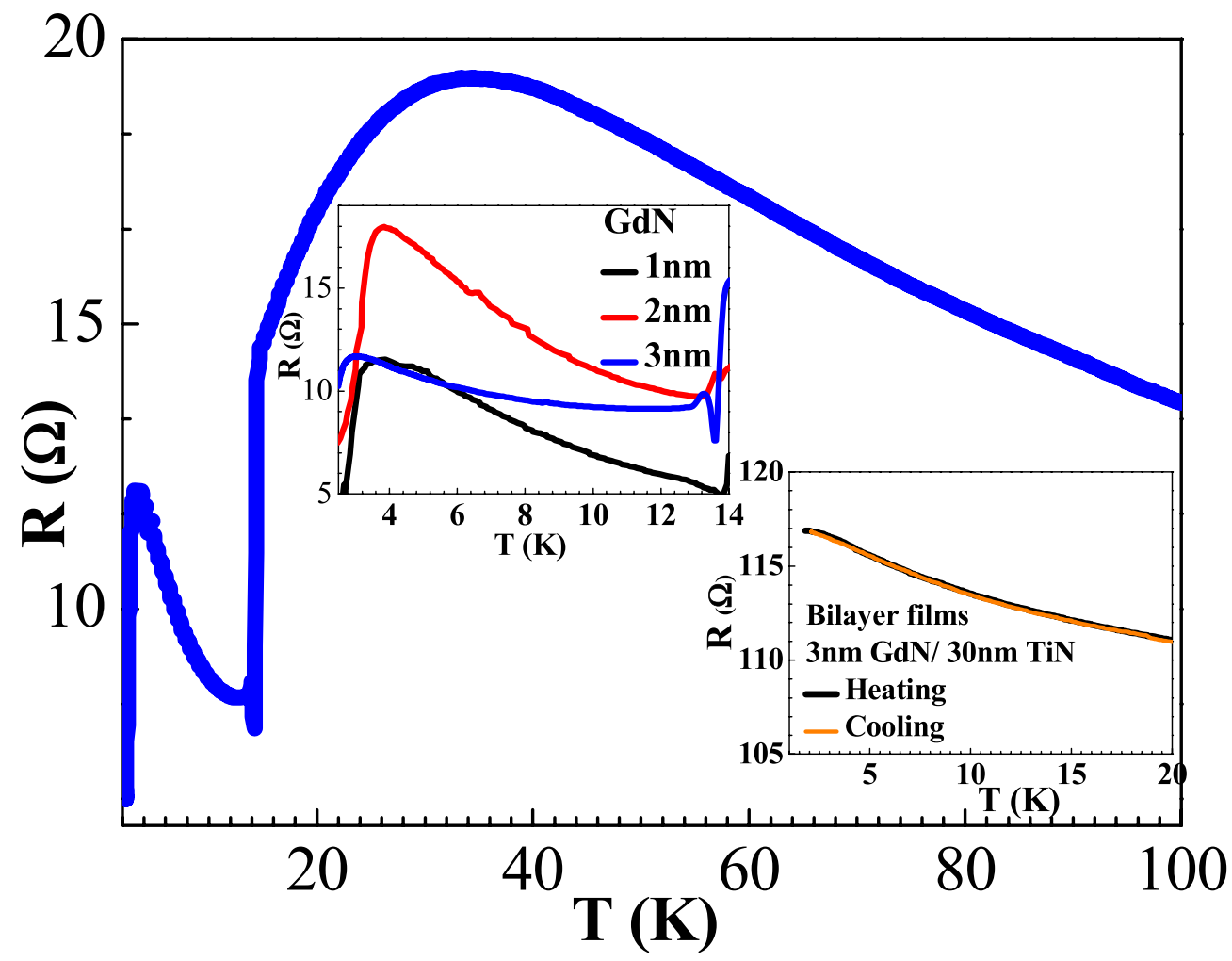

**Figure 1. Temperature dependence of junction resistance.** A 3 nm GdN junction measured at low bias. Top inset shows the Resistance vs Temperature (RT) dependence below superconducting transition of the NbN layer for junctions of 3 thicknesses – 1, 2, 3 nms. Bottom inset shows the RT dependence of a bilayer film of GdN and TiN to demonstrate the absence of superconducting transition in TiN films used in this work.

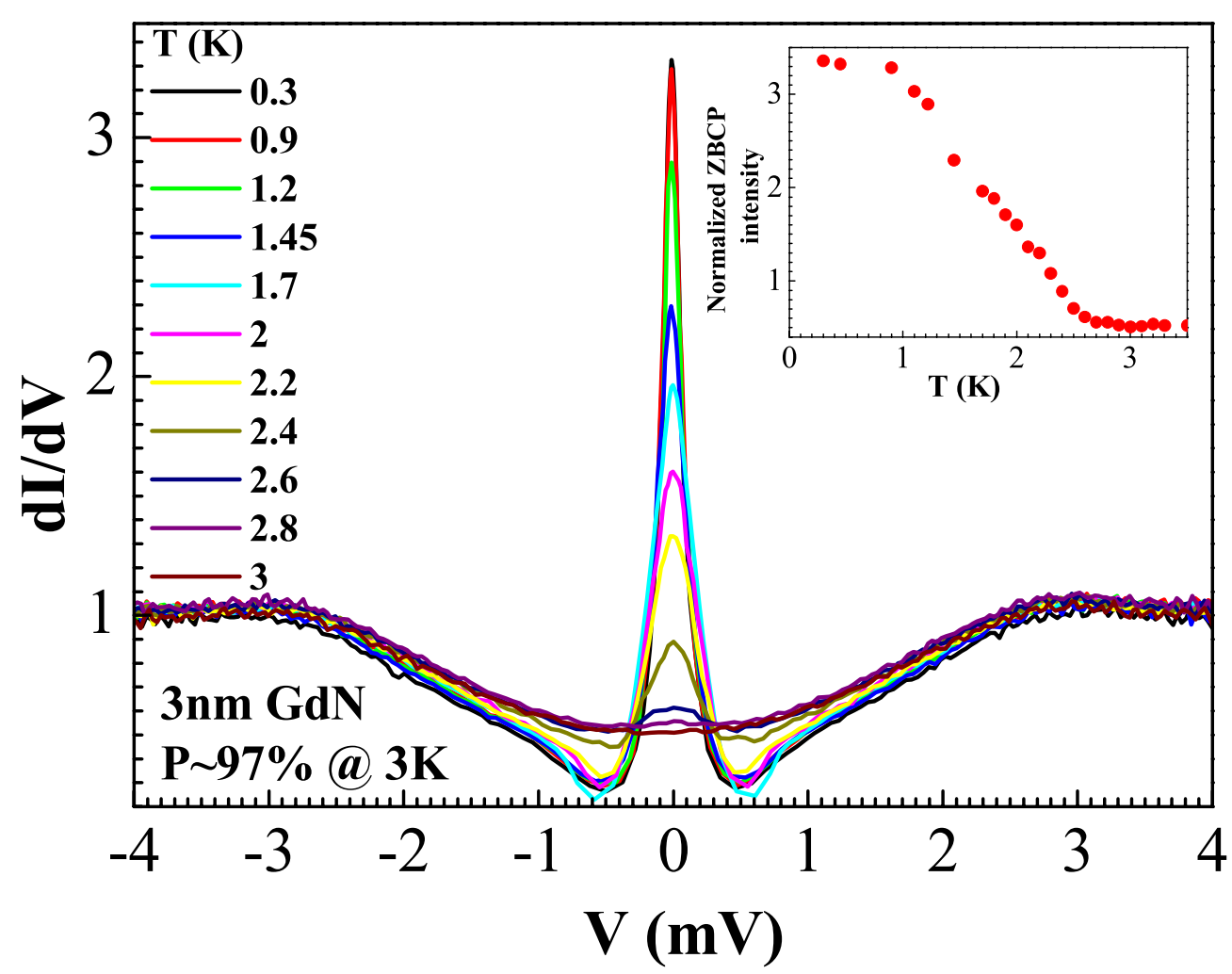

**Figure 2. Evolution of Zero Bias Conductance Peak (ZBCP) with temperature.** Differential conductance (dI/dV) measurements normalised to the normal state conductance of an 100 nm NbN/3 nm GdN/30 nm TiN tunnel junction showing evolution of a ZBCP with decreasing temperature. Inset to the figure shows the temperature dependence of the intensity of the ZBCP.

We observed that a wide range of properties can be obtained in TiN films by altering the nitrogen concentration. In order to obtain non-superconducting TiN, we have tuned the nitrogen concentration (8%) in the sputtering gas mixture.

In Fig. 2, we show the differential conductance curves of a junction with a 3 nm GdN barrier. The curves clearly show the emergence of a strong ZBCP as the junction is cooled to low temperatures. Identical characteristics have been found in all eight junctions on the same chip, and similar characteristics have also been found in all 8 junctions on the same chip of a thinner 2 nm GdN thickness tunnel junction which has spin-polarization

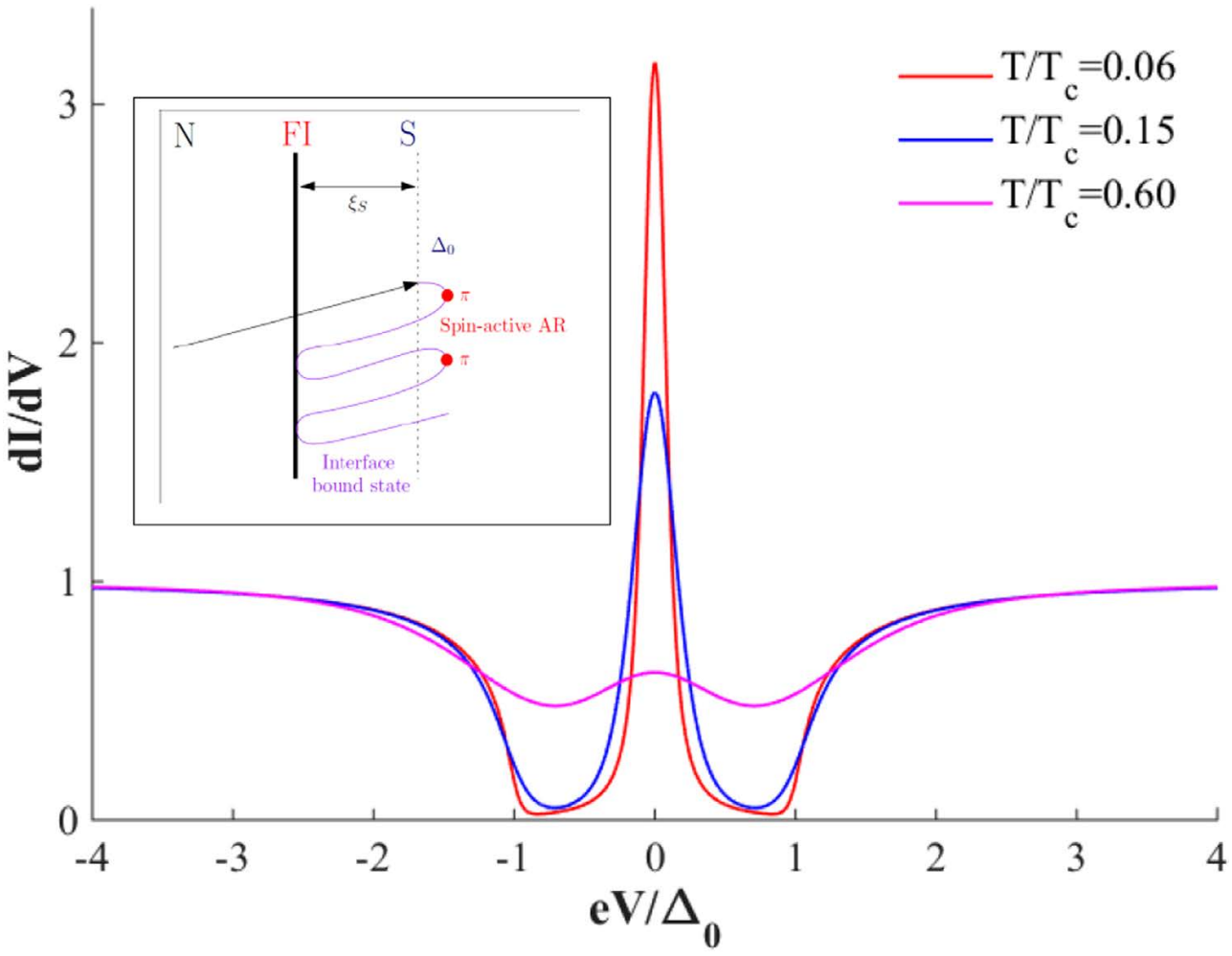

**Figure 3. Theoretical *dI/dV* curves for an S/FI/N junction as a function of applied voltage *eV*.** Following the framework of ref. 22, we have used transmission probabilities $D_\uparrow = 0.20$ and $D_\downarrow = 0.015$ for each spin species, a spin-mixing angle of 0.98π, and set the Dynes parameter to $0.05\Delta_0$. Inset: the formation of a zero-energy bound state at the interface due to spin-active Andreev reflection (AR) by the gap $\Delta_0$ (indicated by red circles) where an additional phase-shift close to π is picked up by the quasiparticles.

(*P*)~65% at 4 K. The ZBCP in all junctions is extremely robust, reproducible, and independent of magnetic field history. The behaviour of these S/I/N junctions at temperatures above which the ZBCP disappears (>3 K), is well understood and has been addressed in detail in a previous publication[17].

It has been theoretically predicted that for spin active interfaces, in the tunnelling limit, a subgap state appears due to spin-dependent phase-shifts[22]. This interface state is manifested via strong conductance peaks at a voltages $eV = \pm\Delta_0\cos(\vartheta/2)$ where $\vartheta$ is the spin-dependent phase shift that is present due to the FI. For $\vartheta = \pi$ the state is pinned to the Fermi level (zero bias). The appearance of this interface state is intimately linked to odd-frequency pairing[13].

ZBCPs are known to occur in several superconducting systems and for a variety of underlying physical mechanisms[4,23–25]. A ZBCP analogous to our experiment observation is the case of d-wave superconductors[24], which occurs due to the sign change of the order parameter at regions in the a–b plane.

For s-wave superconductors, analogous phenomenon can be observed for sign change of the spin dependent phase shift due to the FI which translates to a phase shift of π. Such strong phase shifts[26,27] can be obtained when (a) quasiparticles normal to the interface are the major contributors to the transport process, (b) when spin polarization by the barrier is high, (c) when the barrier profile is not sharp. All the above conditions are met by an NbN/GdN/TiN tunnel junction system, especially that of high spin polarization. An order of magnitude difference between Fermi vectors of NbN[28] and TiN[29] results in quasiparticles normal to the interface being the major contributors to the transport. A previous study has shown that NbN/GdN barrier is different from a conventional box type potential barrier, as a Schottky barrier forms at the NbN/GdN interface[20]. The fact that all the conditions for obtaining a large spin-dependent phase-shift at the interface are met, taken in conjunction with the fact that the conductance spectra demonstrate a ZBCP is a clear indication that this phase-shift is likely to have a value very close to π.

**Theoretical model.** The experimental data in Fig. 2 can be modelled by the theoretical conductance of an S/FI/N structure with a spin-dependent phase-shift close to π, as shown in Fig. 3. The conductance for a ballistic S/FI/N structure has been studied previously[22], and we have followed their analysis when fitting our experimental data. In the tunneling limit, we neglect the suppression of the superconducting order parameter and use the following expression for the current density across the junction

$$J = J_N \int_{-\Delta}^{\Delta} \sum_{\pm} \frac{D_\uparrow D_\downarrow / D}{1 + R_\uparrow R_\downarrow - 2\sqrt{R_\uparrow R_\downarrow}\cos(2\delta \pm \vartheta)} T(E)\,dE + J_N \left( \int_{-\infty}^{-\Delta} + \int_{\Delta}^{\infty} \right) \frac{2\cosh\delta\,[e^{-\delta} D_\uparrow D_\downarrow / D + \sinh\delta]}{e^{2\delta} + e^{-2\delta} R_\uparrow R_\downarrow - 2\sqrt{R_\uparrow R_\downarrow}\cos\vartheta} T(E)\,dE \qquad (1)$$

where $T(E) = \tanh[\beta(E + eV)/2] - \tanh[\beta(E - eV)/2]$.

The following quantities have been defined in Eq. (1): $D = D_\uparrow + D_\downarrow$, $J_N$ is the current density when the superconductor is in its normal state ($J_N \propto D$), $D_\sigma$ and $R_\sigma$ are the probability coefficients for transmission and reflection of spin σ carriers, respectively, $\beta = (k_B T)^{-1}$, $V$ is the applied voltage, $T$ is the temperature, $E$ is the quasiparticle energy, Δ is the superconducting gap, $\vartheta$ is the spin-dependent phase-shift due to the magnetic barrier, and

$$\delta = \text{acos}(E/\Delta) \text{ for } |E| < \Delta,$$
$$\delta = \text{acosh}(|E|/\Delta) \text{ for } |E| > \Delta. \quad (2)$$

For the theoretically simulated conductance plots, we have differentiated Eq. (1) with respect to voltage and normalized the conductance against the normal-state conductance obtained at large voltages $eV \gg \Delta$. To model inelastic scattering, we have incorporated a Dynes parameter via the relation $E \to E + i\Gamma$ where $\Gamma$ provides the quasiparticles with a finite lifetime. The model also accounts for the large difference in tunnelling probability for majority and minority carriers, as expected for a strongly polarized FI.

The temperature-evolution of the conductance spectra matches only qualitatively: the ZBCP vanishes experimentally more rapidly with temperature than in the theory, the reason for this is unclear. However, it must be noted that the temperature dependence of ZBCP is consistent with previous experimental observations of qualitatively similar origins of ZBCP. STM measurements of LDOS in Nb/Ho systems (due to odd frequency triplet superconductivity) observed the ZBCP disappearing at 660 mk[7] – far below the superconducting transition of Nb used in the experiment ($T_{c,Nb}$~6.6 $K$, please refer to supplementary information section of ref. 7), while ZBCPs in YBCO (originating due to sign change of order parameter in d-wave superconductors) were only observed until 40 K and 60 K ($T_{c,YBCO}$~90 $K$) in refs 24 and 30 respectively. We therefore assume that the temperature dependence arises due to aspects of theory which have not been fully understood.

## Conclusions

We have not seen oscillatory behaviour in the intensity of ZBCPs with the application of magnetic field, thus ruling out the possibility of attributing the observed ZBCP to possible Majorana bound states[4]. ZBCPs occurring due to Kondo effects, on application of an external magnetic field, should separate out to a double peak structures[31]. The strong intensity of the ZBCP (3.5 times the normal state conductance) rules out other possibilities like de Gennes-Saint-James resonances[23] or a pin hole mediated junction which in accordance to the BTK theory[32] should give rise to a maximum ZBCP intensity of twice the normal state conductance. ZBCPs could also occur due to the TiN layer turning superconducting thus facilitating a critical current. However, the monotonic field-suppression and the observation of the ZBCP at high magnetic fields clearly indicate that Josephson effects do not cause the ZBCPs. Moreover, the top inset to Fig. 1 – shows that the ZBCPs start to evolve at 2.8 K, 3.8 K and 3.6 K for 3 nm, 2 nm and 1 nm barrier thicknesses respectively. Since the TiN layer for all these films were grown without breaking the vacuum and with the same plasma, this non-monotonic behaviour cannot be related to any possible superconductivity in TiN. However, such temperature dependence again points to an incomplete understanding of theoretical origins for ZBCPs for unconventional superconducting orders. For a more detailed analysis - which rules out superconductivity in TiN layer – please refer to the supplementary information section. Hence, none of the above possibilities are suitable in explaining the observed ZBCPs in our experiment.

The ZBCPs in NbN/GdN/TiN tunnel junctions therefore clearly establish an unconventional non-BCS type DOS indicating odd frequency superconductivity evolving at NbN/GdN interfaces. The current discovery of odd frequency pairing is not only relevant in understanding superconductivity beyond the conventional scope of BCS theory; but also firmly establishes FIs as important material systems for developing active devices for superconducting spintronics[33].

## Methods

The trilayered films of NbN/GdN/TiN are grown without breaking the vacuum in an ultra high vacuum chamber, by means of reactive dc magnetron sputtering in an atmosphere of Argon and Nitrogen. TiN is here grown as a (non-superconducting) metallic layer. Mesa type tunnel junctions were fabricated from sputtered tri-layered films by means of a fabrication procedure described elsewhere[21]. The only difference was that instead of plasma etching, TiN had to be Ar ion milled controllably. Measurements were performed using a 3He dip probe in a closed cycle liquid helium cooled variable temperature insert capable of cooling down to 0.3 K. Spin polarization was calculated from resistance vs temperature measurements using a procedure described in a previous publication[16].

## Acknowledgements
This work of A.P. and M.G.B. was partially supported by an ERC Advanced Investigators Grant 'Superspin'. M.E. and M.G.B. acknowledges financial support from the EPSRC grant EP/N017242/1. M.E. also acknowledges funding from the Lars Onsager committee during his stay as Lars Onsager Professor at NTNU. J.A.O. and J.L. acknowledge funding via the "Outstanding Academic Fellows" programme at NTNU, the COST Action MP-1201 and the Research Council of Norway Grant numbers 205591, 216700, and 240806. The authors acknowledge helpful discussions with Jason Robinson.

## Author Contributions
M.G.B. and A.P. designed the experiment. A.P. grew the films, fabricated the devices and performed the measurements. J.A.O., M.E. and J.L. did the theoretical modelling. All authors contributed equally to writing the manuscript.

## Additional Information
**Supplementary information** accompanies this paper at http://www.nature.com/srep

**Competing financial interests:** The authors declare no competing financial interests.

# Supplementary information

## Spectroscopic evidence of odd frequency superconducting order

Avradeep Pal[1*], J. A. Ouassou[2], M. Eschrig[3], J. Linder[2*], M. G. Blamire[1]

[1]*Department of Materials Science, University of Cambridge, 27 Charles Babbage Road, Cambridge CB3 0FS, United Kingdom*
[2]*Department of Physics, Norwegian University of Science and Technology, N-7491 Trondheim, Norway*
[3]*SEPnet and Hubbard Theory Consortium, Department of Physics, Royal Holloway, University of London, Egham, Surrey TW20 0EX, United Kingdom*

We provide the following evidence to rule out superconductivity in the TiN layer and establish conclusively that the ZBCP is due to SIN junctions and therefore represents an enhancement in DOS around the Fermi level in NbN:

### a) Variation of gap edge with temperature

If TiN turned superconducting at 3K ($T_{c\,TiN}$~$3K$), and assuming that TiN behaves like an ideal BCS superconductor, $\Delta_{TiN}\,(0)$~$1.76\;k_B T_{c\,TiN} = 0.455mV$.

At 1.6K: $\Delta_{TiN}\,(1.6K)$~$\Delta_{TiN}\,(0)*\tanh\left(1.74*\sqrt{\left(\frac{T_{c\,TiN}}{1.6}-1\right)}\right)$ = 0.42mV.

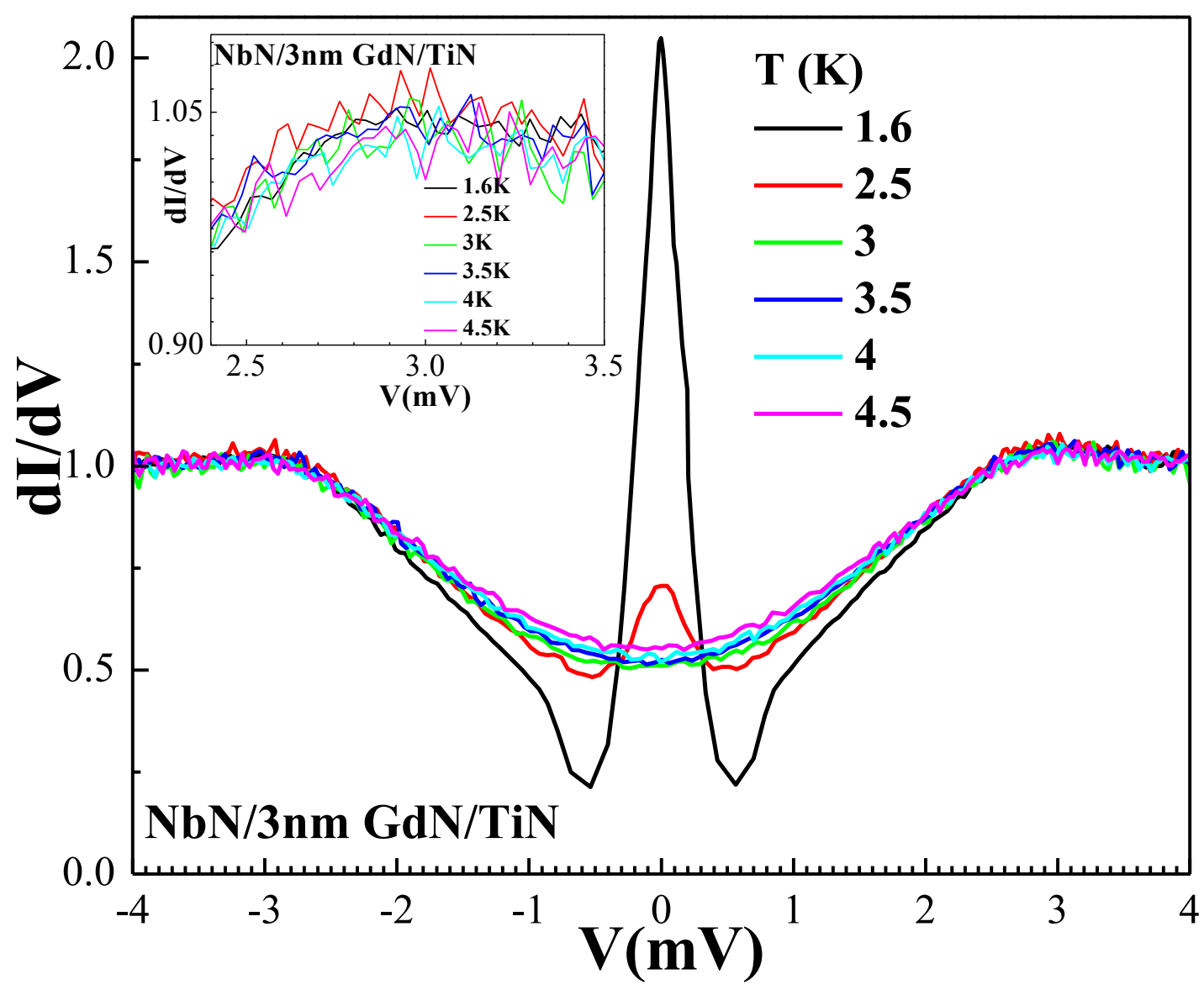

Supplementary Figure 1 – Differential conductance curves at temperatures from 4.5K to 1.6K demonstrating that there is no observation in enhancement of gap edge due to possible superconductivity of TiN layer.

Hence we would expect the gap edge in NbN/GdN/TiN junctions to increase by ~ 0.5 mV as the junction is cooled. The figure below and the inset to it, clearly indicates that no such increase is observed. In fact, no increase is observed even at 0.3K (Refer to Fig. 2 of the main manuscript). This is a clear demonstration that NbN is the only superconducting layer in the tri-layer stack. Owing to high transition temperature of NbN ($T_{c\,NbN}$~13.5K), the magnitude of its gap edge feature saturates below approx. 7K. Thus, these junctions are SIN junctions throughout the entire temperature range below 13.5K.

**b) <u>Discussion on a potential Josephson effect</u>**

Josephson junctions with two superconducting NbN electrodes [1] with exactly identical junction dimensions to the devices reported here exhibit $I_c(H)$ Frauhofer patterns with an oscillation period or lobe width of of 1.5-3.5 mT. If we assume that, despite the arguments above, that the TiN is superconducting then we can take its penetration depth ($\lambda_L$) to be $\sim 700nm$, and therefore should have lobe width:

$$H_0 = \Phi_0/(L\left(\lambda_{NbN}{}^*\tanh\left(\frac{d_{NbN}}{\lambda_{NbN}}\right) + \lambda_{TiN}{}^*\tanh\left(\frac{d_{TiN}}{\lambda_{TiN}}\right) + d_{GdN}\right))$$

where $d_{NbN} = 100nm, d_{TiN} = 30nm, d_{GdN} = 3nm,\ \lambda_{NbN} \sim 250nm$ and $\Phi_0$ is the universal flux quantum. As per the above expression – the Fraunhofer patterns should have critical current oscillation periods as approx. 4.4mT. Taking the standard decay envelope of the Fraunhofer pattern, we can therefore say that any Josephson effects should die out at high fields (>45mT; i.e. – beyond the supposed $10^{th}$ Fraunhofer lobe). However, the ZBCPs in our samples are only suppressed slowly and monotonically and remain visible even till 1 T, thus ruling out ZBCPs due to critical currents (conductance maxima) at low bias due to Josephson effects.

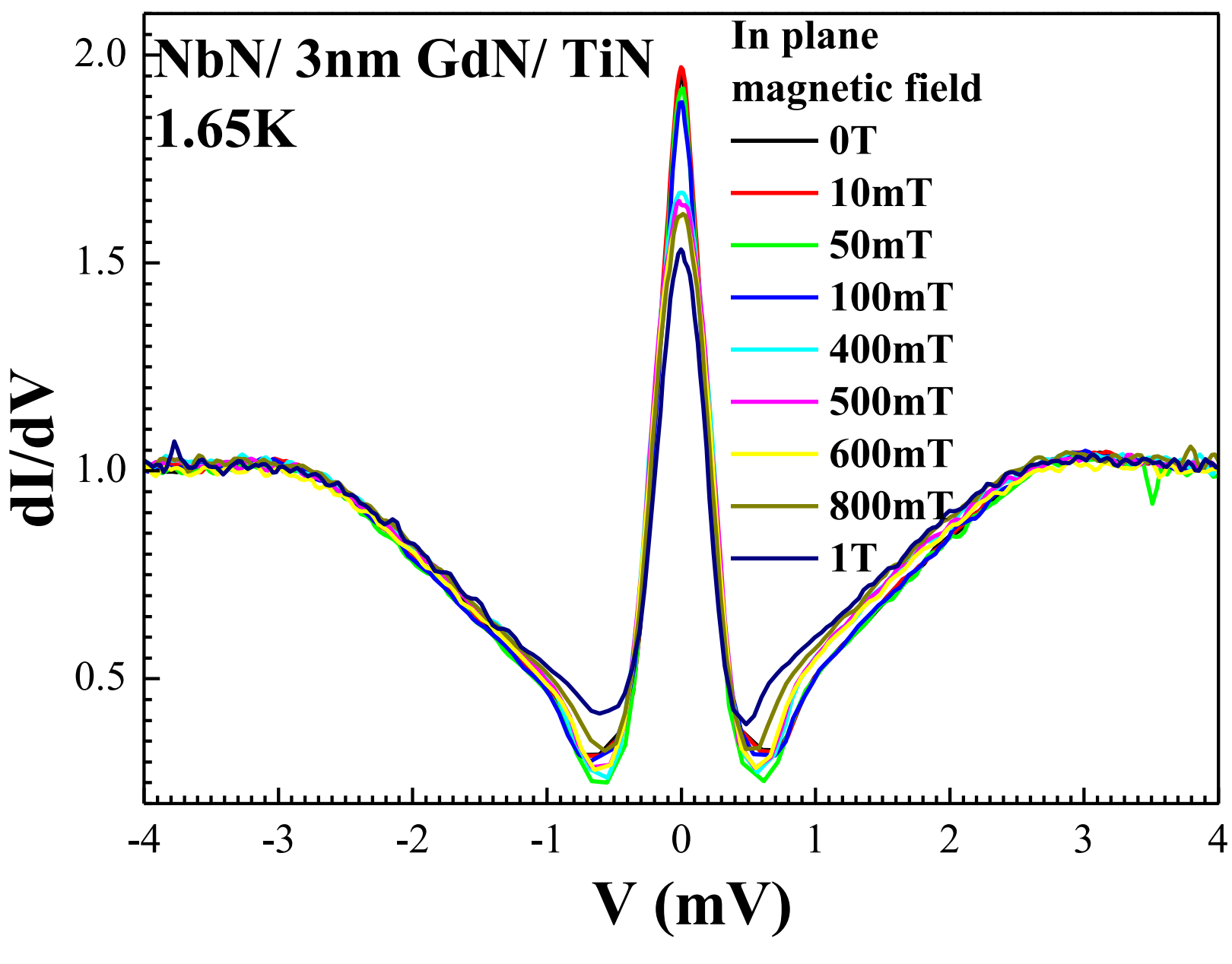

Supplementary Figure 2 – Differential conductance measurements of the ZBCP at different values of externally applied in-plane magnetic fields.

c) **<u>Comparison of IV curves of SIS and SIN junctions</u>**

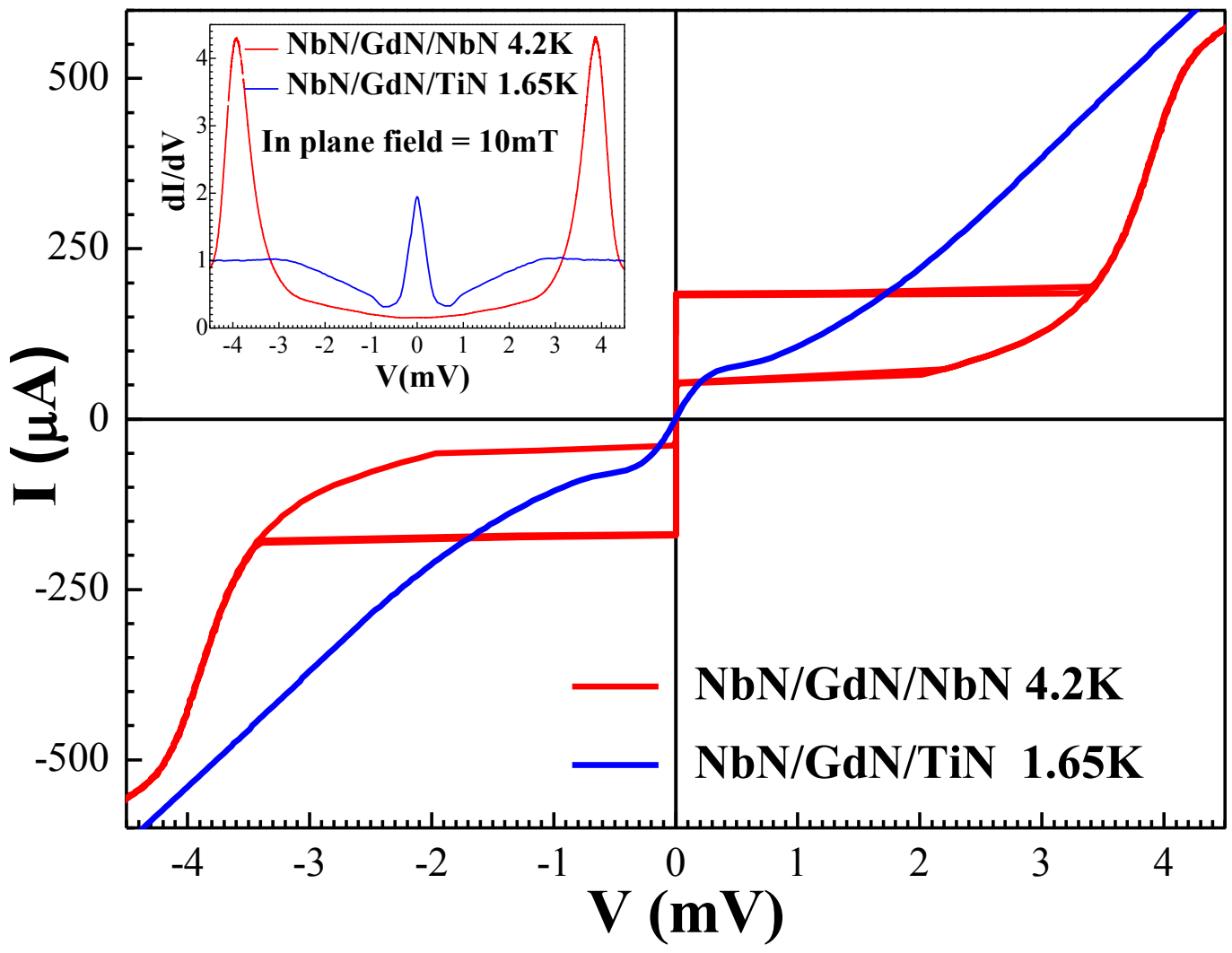

Supplementary Figure 3 – IV curves (including forward and backward traces) of two junctions, one with a superconducting NbN counter-electrode (red), while the other has non-superconducting TiN electrode (blue). IV curves are measured at zero field and well below the supposed superconducting Tc of NbN (13.5K) and TiN (either 5.4K or 3K as suggested by referee). Inset to figure shows the comparison of respective differential conductance curves measured with an in-plane magnetic field of 10mT.

The most noticeable features in Figure 3 are that for the (SIN) junction with the TiN counter-electrode (blue curve), the high conductance region around zero bias is not vertical and that switching away from this region (which occurs at roughly the same magnitude of current ($\sim \pm 65\mu A$) as the critical current of the NbN/GdN/NbN SIS junction) shows no sign of hysteresis – a feature that is expected in underdamped SIS Josephson junctions and clearly present in the NbN/GdN/NbN device. However, for the junction with NbN counter-electrode, we see clear hysteresis (Switching current $\sim \pm 168\mu A$, Re-trapping current $\sim \pm 37\mu A$). We have reported in a prior publication [2] that our GdN junctions are in the underdamped regime, and hence the above IV traces clearly demonstrate that the high conductance region in TiN junctions is not a feature caused by critical currents, but rather, represents an enhancement in the DOS around the Fermi level of the NbN superconductor for the NbN/GdN/TiN device.

The inset shows the previously discussed effect of magnetic field on the supercurrent due to the Josephson Effect (Fraunhofer pattern). While the critical current in NbN/GdN/NbN junction is completely suppressed due to the 10 mT in-plane magnetic field; an identical in-plane field has no effect on the magnitude of the ZBCP.